\documentclass{iau}
\usepackage{graphicx}

\title[Pulsation of magnetic stars] 
{Pulsation of magnetic stasrs}

\author[Hideyuki Saio]   
{Hideyuki Saio}

\affiliation{Astronomical Institute, Graduate School of Science, Tohoku University, 
\\ Aoba-ku, Sendai, Japan
\\ email: {\tt saio@astr.tohoku.ac.jp}}

\pubyear{2013}
\volume{301}  
\pagerange{1--8}
\jname{Precision Asteroseismology}
\editors{J.A. Guzik, W.J. Chaplin, G. Handler \& A. Pigulski, eds.}
\begin{document}

\maketitle

\begin{abstract}
Some Ap stars with strong magnetic fields pulsate in high-order p modes; they are called roAp (rapidly oscillating Ap) stars.
The p-mode frequencies are modified by the magnetic fields. 
Although the large frequency separation is hardly affected, 
small separations are modified considerably. 
The magnetic field also affects the latitudinal amplitude distribution on the surface. 
We discuss the property of axisymmetric 
p-mode oscillations in roAp stars.
\end{abstract}

\def\bm#1{\mbox{\boldmath$#1$}}

\firstsection 
\section{Observational properties of roAp stars}

\begin{figure}[b]
\vspace*{-0.2 cm}
\begin{center}
\includegraphics[width=0.5\textwidth]{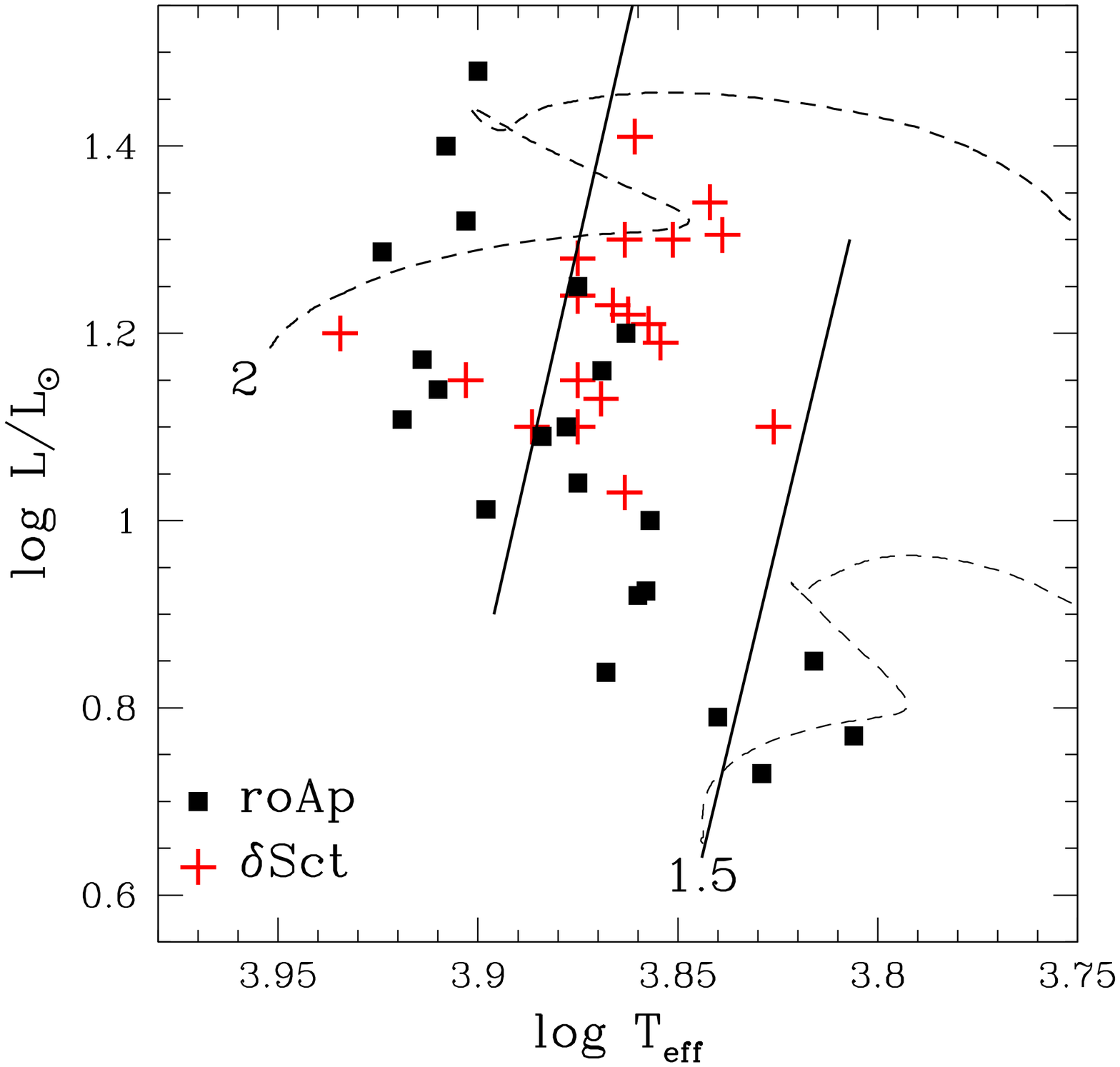} 
 \hspace{-0.04\textwidth}
 \includegraphics[width=0.5\textwidth]{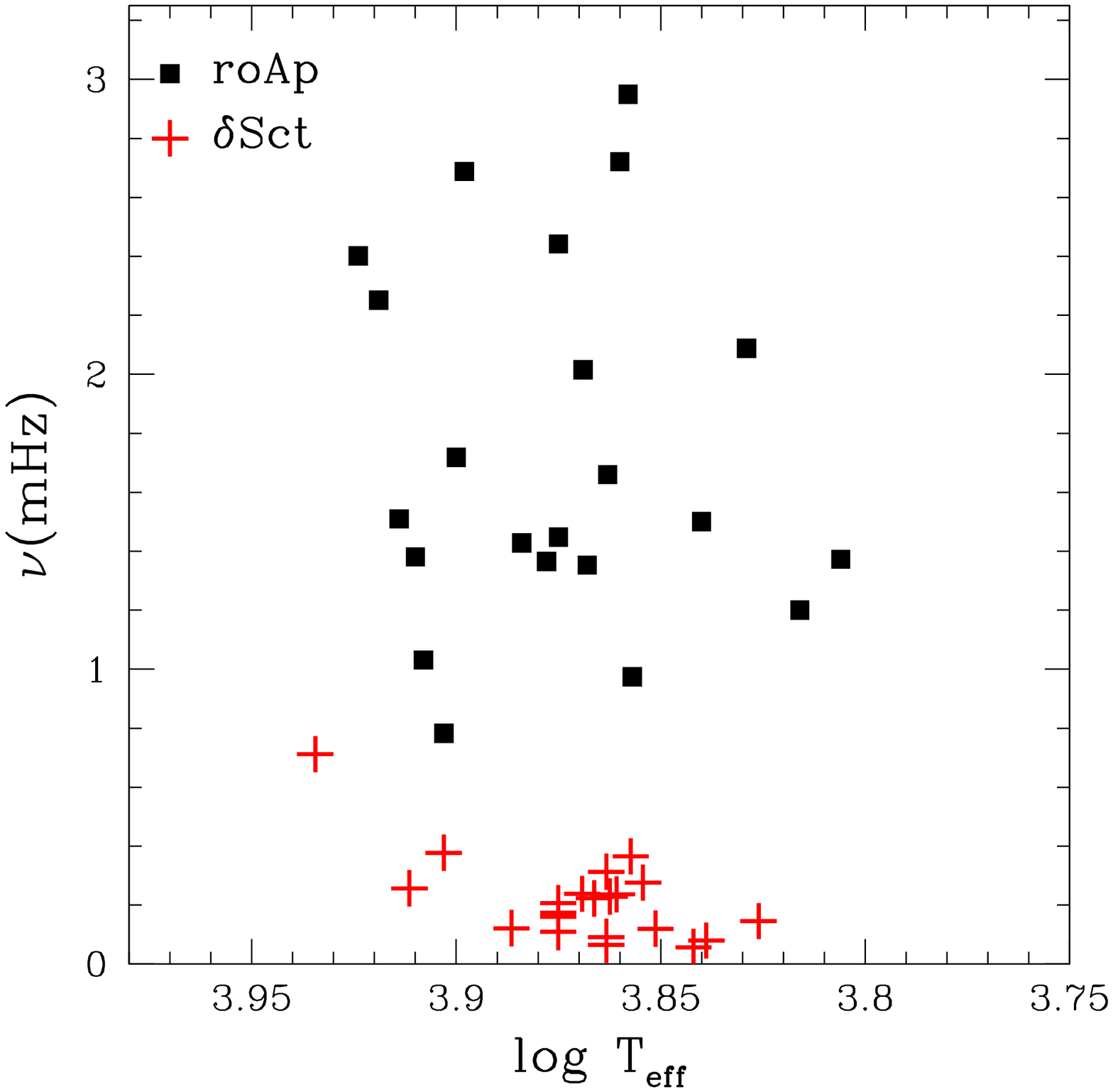} 
 \vspace*{-0.2 cm}
\caption{\textit{Left:} The positions of roAp stars (filled squares) in the H-R diagram compared with the positions $\delta$ Sct stars (pluses).  Two solid lines indicate the stability boundaries of radial fundamental modes obtained by \cite[Dupret et al.~(2005)]{dup05} including time-dependent convection.
\textit{Right:} Main pulsation frequencies of the stars in the left panel are plotted with respect to the effective temperature. 
}
\label{fig:hrd_freq}
\end{center}
\end{figure}

The  group of rapidly oscillating Ap (roAp) stars was discovered by \cite[Kurtz (1982)]{kur82} more than thirty years ago (it consisted of five members then). 
Since then, the number of known roAp stars has increased to around 40--50  (still increasing); a recent list is given by \cite[Kurtz et al.~(2006)]{kur06}. 
They are chemically-peculiar main-sequence stars having  masses ranging from $\sim 1.5$ to $\sim 2\,M_\odot$, and effective temperatures from $\sim 6400$ to $\sim 8400$\,K.
On the H-R diagram (Fig\,\ref{fig:hrd_freq}, left panel), they lie in and around the $\delta$ Sct instability strip.  
However, the oscillation frequencies are much higher than those of $\delta$ Sct variables as seen in the right panel of Fig.\,\ref{fig:hrd_freq}, which plots dominant frequencies of roAp and $\delta$ Sct stars with respect to the effective temperature. 
The roAp stars pulsate in high-order p modes whose frequencies range from $\sim 1$ to $\sim 3$\,mHz (periods; $\sim 6 - 21$\,min), while $\delta$ Sct stars pulsate in low-order p modes. The pulsations are generally multi-periodic, and some stars show nearly equally spaced frequencies; e.g., HR 1217 (\cite[Kurtz et al.~2005]{kur05}).

\begin{figure}[t]
\begin{center}
 \includegraphics[width=0.5\textwidth]{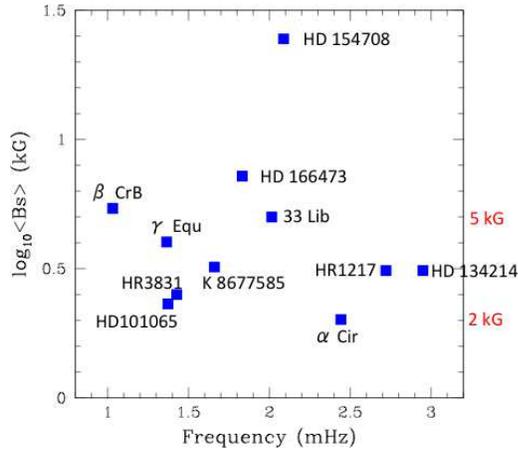} 
 \caption{Surface magnetic field moduli of some roAp stars are plotted with respect to the dominant oscillation frequencies. 
The magnetic field strengths were taken from \cite[Shulyak et al.~(2013)]{shu13}, \cite[Ryabchikova et al.~(2008)]{rya08}, \cite[Mathys et al.~(1997)]{mat97}, \cite[Hubrig et al.~(2005)]{hub05}, \cite[Kochukhov et al.~(2004)]{koc04}, \cite[Balona et al.~(2013)]{bal13}.}
   \label{fig:magfields}
\end{center}
\end{figure}

Another important property of roAp stars is the presence of strong global (mainly dipole) magnetic fields.
Fig.\,\ref{fig:magfields} shows the surface magnetic field moduli of some roAp stars as a function of the dominant pulsation frequency.
The field strength ranges from $\sim 1$ to 24\,kG (mostly between 2 and 6\,kG), which shows no appreciable correlation with the pulsation frequency.
The lack of a correlation might be indicating that the magnetic field itself does not play a direct role in exciting high-order p modes in roAp stars.

\section{Excitation of  high-order p modes in roAp stars}

\begin{figure}[t]
\begin{center}
 \includegraphics[width=0.55\textwidth]{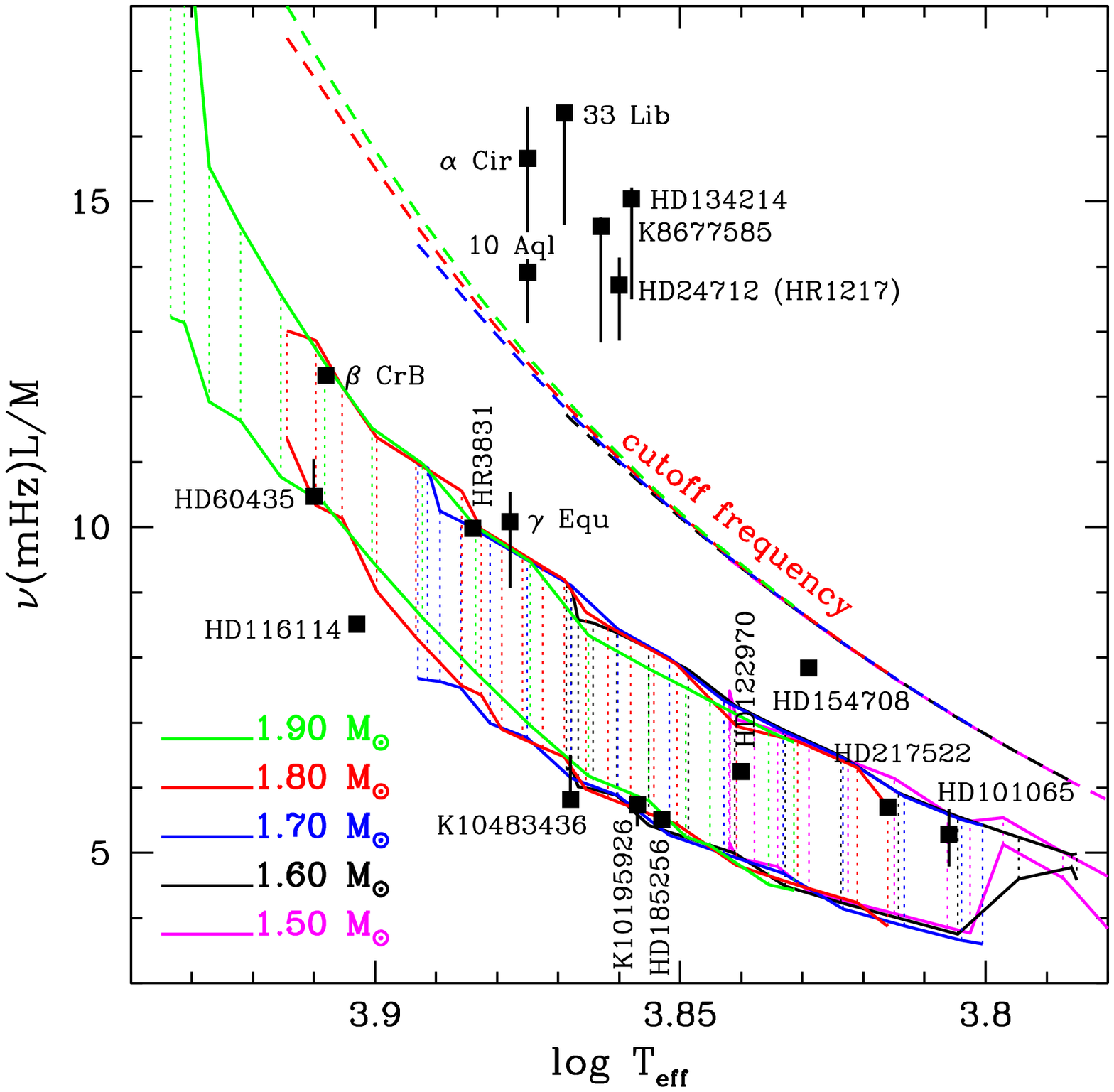} 
\vspace*{-0.5 cm}
 \caption{The instability region of high-order p modes (shaded area) in the $\log T_{\rm eff}-\nu L/M$ plane, where $\nu$ is oscillation frequency, and luminosity $L$ and mass $M$ are in solar units. Also plotted are some of the roAp stars whose parameters are relatively well known. Dashed lines are acoustic cutoff frequencies along evolutionary tracks of models with various masses.}
   \label{fig:stability}
\end{center}
\end{figure}

The $\kappa$ mechanism in the He II ionization zone, which excites low-order p modes in $\delta$~Sct stars, is ineffective in exciting high-order p modes in roAp stars, because the thermal time there is much longer than the periods.   
Rather, the H ionization zone is more important for roAp stars, although the convective flux weakens the effect of the $\kappa$ mechanism.
\cite[Balmforth et al.~(2001)]{bal01} found that if a strong magnetic field suppresses the convection in the H ionization zone in the polar regions, high-order p modes are excited.
Based on this assumption, \cite[Cunha (2002)]{cun02} obtained an instability region for high-order p modes, which is largely consistent with the positions of roAp stars in the H-R diagram.  

The comparison on the H-R diagram, however, does not tell whether the mechanism works for the frequencies of actual roAp stars (Fig.\,\ref{fig:hrd_freq}).
A better comparison might be possible on
the $\log T_{\rm eff}-\nu L/M$ plane ($\nu=$~oscillation frequency).
In this plane,  the minimum and maximum frequencies of excited high-order p modes at various main-sequence evolution stages of various masses are located, respectively, on two lines, forming an instability range.
Fig.\,\ref{fig:stability} shows the instability region and the positions of some roAp stars whose parameters are relatively well determined.
Obviously, some of the well studied roAp stars are far from the instability region, and their oscillation frequencies seem to be above the acoustic cutoff frequency.
In fact, pulsation phase variations in the atmosphere of  some of these stars indicate the presence of running waves (e.g., HR 1217, \cite[Saio et al.~2010]{sai10}; HD 134214, \cite[Saio et al.~2012]{sai12}). 
We still do not fully understand the excitation of  high-order p modes in roAp stars. 

\section{Effect of magnetic fields on high-order p-mode pulsations}
\subsection{Linearized basic equations}
\begin{figure}[t]
\begin{center}
 \includegraphics[width=0.5\textwidth]{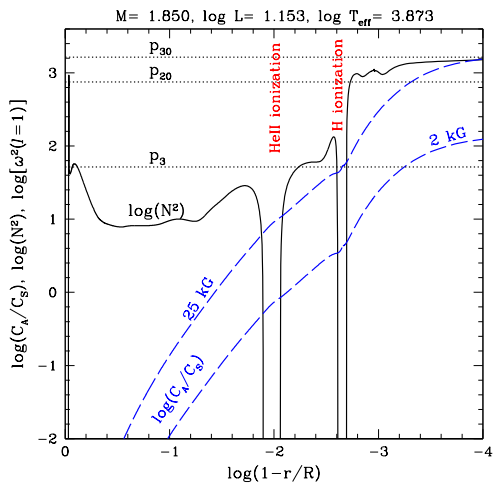} 
\caption{Runs of Brunt-V\"ais\"al\"a frequency, $N$, and the ratio of the Alfv\'en speed to the sound speed, $C_{\rm A}/C_{\rm S}$, for magnetic field strengths of 2 and 25\,kG. Horizontal dotted lines indicate the square of pulsation frequencies of some dipole p modes.  The squares of frequencies, $\omega^2$ and $N^2$ are normalized by $GM/R^3$, with $G$ being the gravitational constant. }
   \label{fig:n2}
\end{center}
\end{figure}

In the presence of a magnetic field, the motion of the ionized gas is affected by the Lorentz force $\bm{(\nabla\times B)\times B/(4\pi \rho)}$ (in the MHD approximation with cgs units) with $\rho$ being the gas density. 
The ratio of the Lorentz force to the pressure-gradient force, $|(\nabla p)/\rho|$, is estimated roughly as $(C_{\rm A}/C_{\rm S})^2$, where the Alfv\'en speed $C_{\rm A}$ and sound speed $C_{\rm S}$ are defined as
\begin{equation}
C_{\rm  A} = {B\over \sqrt{4\pi\rho}} \quad \mbox{and} \quad  C_{\rm S}=\sqrt{\Gamma_1{p\over\rho}}
\end{equation}
where $\Gamma_1$ is the first adiabatic index. 
The runs of the ratio $C_{\rm A}/C_{\rm S}$ for $B=2$ and 25\,kG shown in Fig.\,\ref{fig:n2} indicate that the Lorentz force is important in the layers as deep as the He~II ionization zone in the roAp stars.  
Geometrically, however, the layers are thinner than $\sim5$\% of the stellar radius. 

Including the Lorentz force, a linearized momentum equation for nonradial pulsations may be written as
\begin{equation}
{d\bm{v}\over dt} = {\rho'\over\rho^2}{dp\over dr}\bm{e}_r 
- {1\over\rho}\nabla p' +{1\over 4\pi\rho}(\nabla\times\bm{B}')\times\bm{B}_0,
\label{eq:moment}
\end{equation}
where $\bm{v}$ is pulsation velocity,  and $\bm{e}_r$ is the unit vector in the radial direction. The prime $(')$ indicates the Eulerian perturbation of the quantity, and the subscript $0$ on $\bm{B}$ means its equilibrium value. The equilibrium magnetic field is assumed to be force free ($\nabla\times\bm{B}_0=0$; a dipole field is assumed). 
We have used the Cowling approximation, in which the Eulerian perturbation of  gravitational potential is neglected.

Assuming the ideal MHD condition (neglecting magnetic diffusivity), the linearized magnetic induction equation is given as
\begin{equation}
{\partial\bm{B}'\over\partial t}= \nabla\times(\bm{v}\times\bm{B}_0).
\label{eq:induction}
\end{equation}
In addition to these equations, the linearized mass-conservation equation
\begin{equation}
{\partial\rho'\over\partial t} + \nabla\cdot(\rho\bm{v}) = 0,
\label{eq:cont}
\end{equation}
and the adiabatic relation $\delta p/p = \Gamma_1\delta\rho/\rho$ form a closed set of equations for adiabatic oscillations, where $\delta$ means the Lagrangian perturbation.

\subsection{Local analysis -- Magneto-acoustic waves}
To understand the basic properties of oscillations, we use here a local analysis, where perturbations are assumed to be proportional to $\exp(i\sigma t - ik_zz-ik_xx)$ 
with the plane-parallel approximation; $z$ and $x$ are distances in radial and latitudinal directions. 
This form corresponds to axisymmetric modes, in which the Alfv\'en wave (torsional) is excluded. 
Substituting this form of perturbation into eqs. (\ref{eq:moment}) -- (\ref{eq:cont}) and using the adiabatic relation we have a dispersion relation,
\begin{equation}
\sigma^4-\sigma^2[N^2+k^2(C_{\rm S}^2+C_{\rm A}^2)] + C_{\rm S}^2C_{\rm A}^2k^2k_\shortparallel^2+N^2\left(C_{\rm S}^2k_x^2+C_{\rm A}^2k^2{B_z^2\over B^2}\right)=0
\label{eq:disp}
\end{equation} 
(Appendix A in \cite[Saio \& Gautschy 2004a]{sai04a}), where $k^2=k_z^2+k_x^2$, $k_\shortparallel$ is the wavenumber parallel to the magnetic field (i.e., $k_\shortparallel=\bm{k}\cdot\bm{B}/|B|$), and $N$ is the Brunt-V\"ais\"al\"a frequency. 

Since, in most of the cavity of high-order p modes (Fig.\,\ref{fig:n2}), $N^2$ is not important, we neglect $N^2$ in eq.~(\ref{eq:disp}). Solving the equation for $\sigma^2$ we obtain
\begin{equation}
\sigma^2={1\over2}k^2(C_{\rm S}^2+C_{\rm A}^2)\left(1\pm\sqrt{1-{4V_{\rm C}^2\over C_{\rm S}^2+C_{\rm A}^2}{k_\shortparallel^2\over k^2}}\right),
\label{eq:disp_sol}
\end{equation} 
where $V_{\rm C}$ (cusp velocity) is defined as
\begin{equation}
V_{\rm C}^2 = {C_{\rm S}^2C_{\rm A}^2\over C_{\rm S}^2+C_{\rm A}^2}.
\label{eq:cusp}
\end{equation}
If either $C_{\rm S}\gg C_{\rm A}$ or $C_{\rm S}\ll C_{\rm A}$,  $4V_{\rm C}^2/(C_{\rm S}^2+C_{\rm A}^2) \ll 1$.  Then, from eq.~(\ref{eq:disp_sol}) we obtain two types of waves
\begin{equation}
\sigma^2\approx k^2(C_{\rm S}^2+C_{\rm A}^2) \quad \mbox{(fast wave)}, \qquad
\sigma^2\approx k_\shortparallel^2 V_{\rm C}^2 \quad \mbox{(slow wave)}.
\end{equation}
In these conditions, in which $(C_{\rm S}^2+C_{\rm A}^2) \gg V_{\rm C}^2$, fast waves decouple from slow waves, because the wave number of fast waves is much smaller than that of slow waves for a given frequency.
In the condition of $C_{\rm S} \sim C_{\rm A}$, however, the two waves have similar wave numbers and a coupling occurs. 

In the outermost layers where $C_{\rm A} \gg C_{\rm S} \sim V_{\rm C}$ (Fig.\,\ref{fig:n2}, eq.\,\ref{eq:cusp}), 
slow waves correspond to p-mode acoustic oscillations, while
in the deep interior where  $C_{\rm S} \gg C_{\rm A} \sim V_{\rm C}$, fast waves correspond to p-mode pulsations.
In these layers p-mode pulsations decouple from magnetic waves.
In the intermediate layers, in which $C_{\rm S} \sim C_{\rm A} \sim V_{\rm C}$,
the coupling occurs; i.e., p-mode pulsation generates a slow wave which propagates downwards.
\cite[Roberts \& Soward (1983)]{rob83} argued that the slow wave will be dissipated before it reaches the stellar center because the wavelength of the slow wave decreases rapidly as the ratio $C_{\rm A}/C_{\rm S}$ decreases. 
This means that the slow wave carries away a fraction of the pulsation energy, so that the pulsation damps even in the adiabatic analysis. 
This damping is incorporated in later works solving pulsation equations for magnetized stars.

\subsection{Comparison of frequency shifts calculated by different methods}

\begin{figure}[b]
\vspace*{-0.2cm}
\begin{center}
\includegraphics[width=0.52\textwidth]{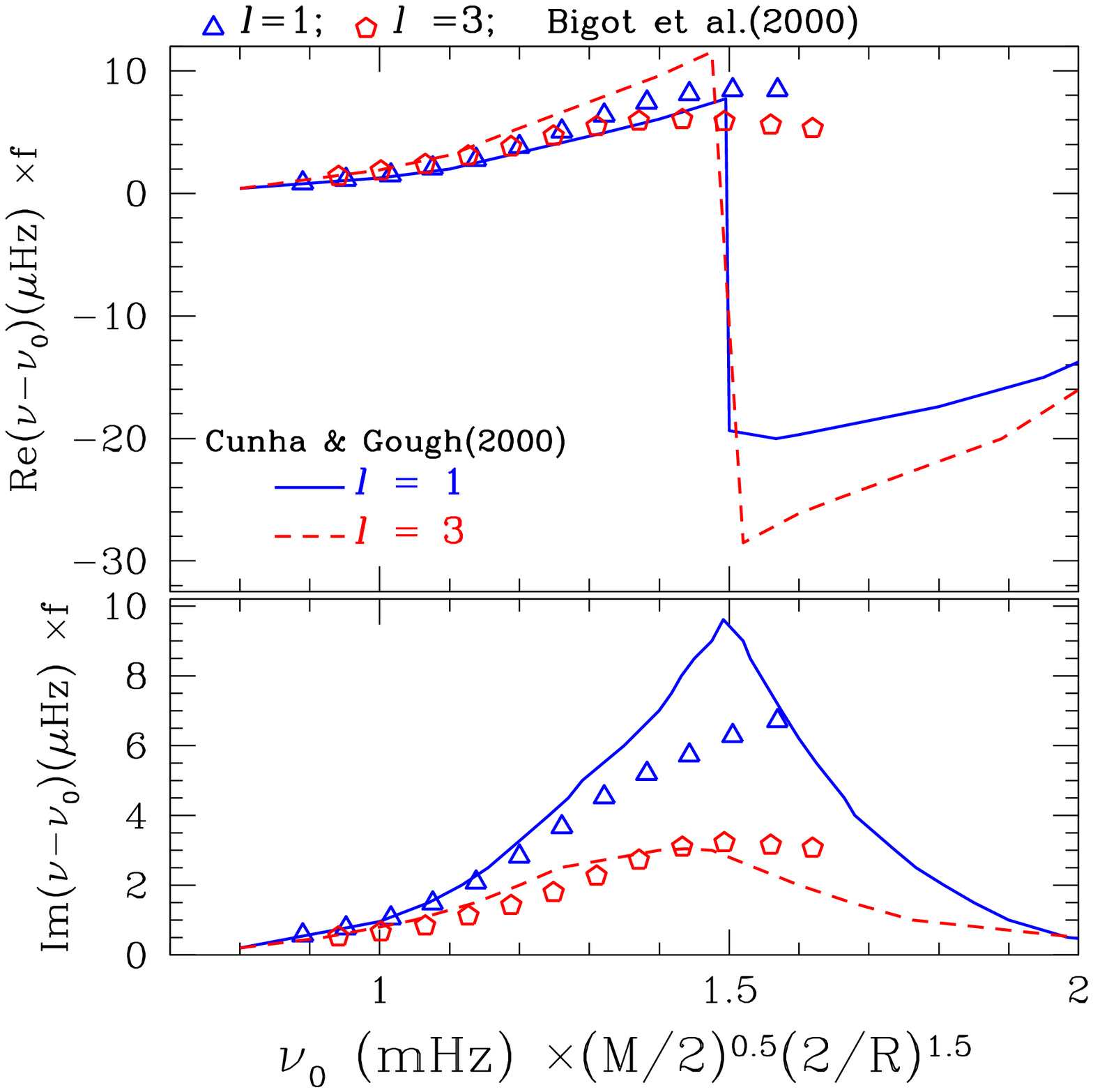}
\hspace{-0.06\textwidth}
\includegraphics[width=0.52\textwidth]{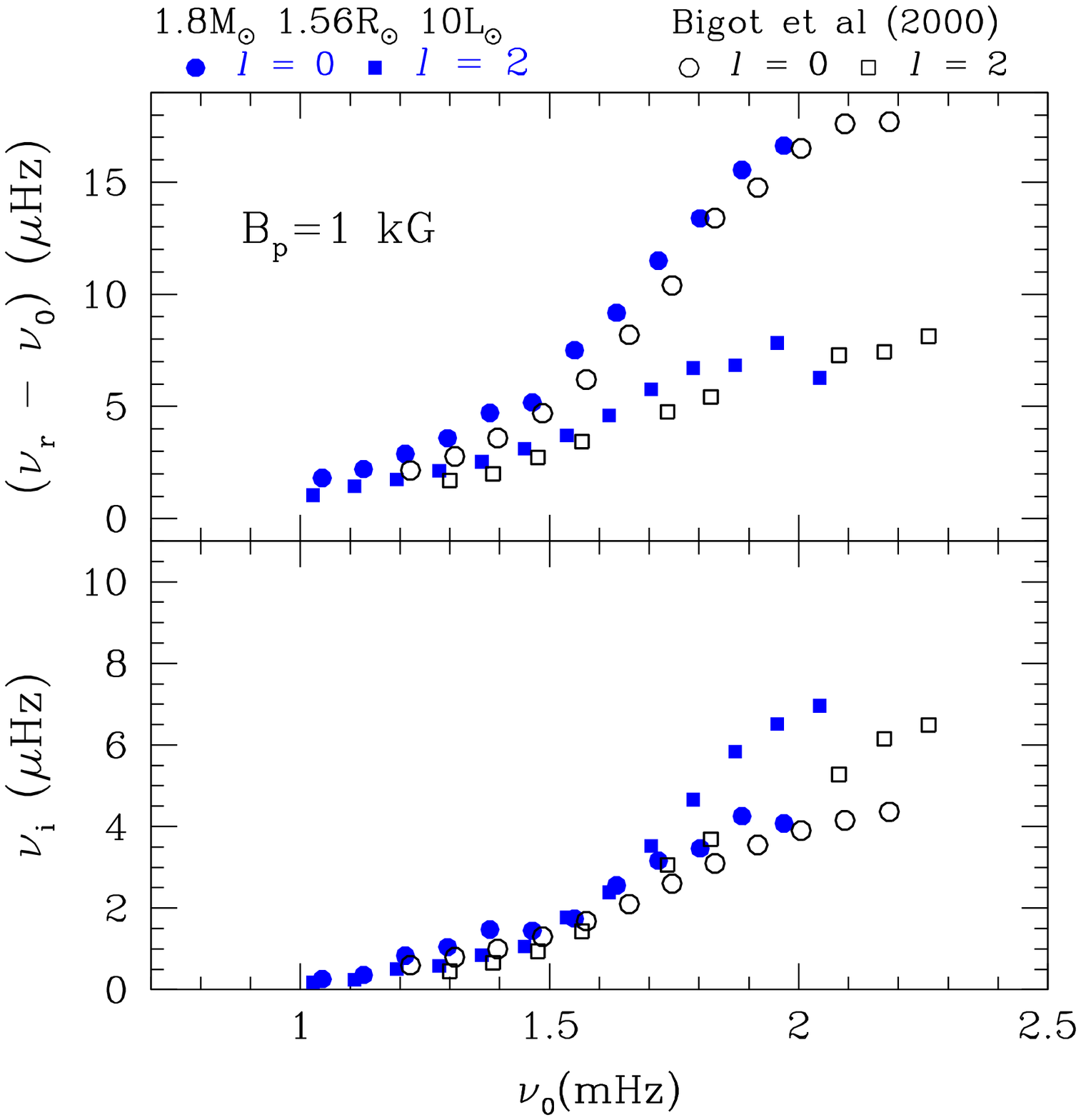}
\vspace{-0.7cm}
\caption{\textit{Left:} Magnetic frequency shifts of high-order p modes at $B_{\rm p}=1$\,kG obtained by \cite[Cunha \& Gough (2000)]{cun00} (solid lines for $\ell=1$ and dashed lines for $\ell=3$) for a $2\,M_\odot$ (and $R=2R_\odot$) polytropic (index 3)  model are compared with the results for a $1.8\,M_\odot$ ZAMS model obtained by \cite[Bigot et al.~(2000)]{big00}. To normalize the difference in mass and radius, a factor
$f= (M/2)^{0.5}(2/R)^{1.5}$ is multiplied to the frequencies. 
\textit{Right:} Comparison
between the results obtained by \cite[Bigot et al.~(2000)]{big00} and by the method of \cite[Saio \& Gautschy (2004a)]{sai04a} for even modes ($\ell=0,2$). }
\label{fig:comparison}
\end{center}
\end{figure}

Three different methods have been developed to calculate oscillation frequencies taking into account the generation of the slow wave and its damping effect; the first one is by \cite[Dziembowski \& Goode (1996)]{dzi96}, which is extended by \cite[Bigot et al.~(2000)]{big00}; the next one is based on a variational principle developed by \cite[Cunha \& Gough (2000)]{cun00}; and the third one  (\cite[Saio \& Gautschy 2004a]{sai04a}) is somewhat similar to the first one, expanding the eigenfunction by a sum of terms associated with spherical harmonics.
These three methods are summarized by \cite[Saio (2008)]{sai08}.

Figure\,\ref{fig:comparison} compares results obtained by the three different methods.
Plotted are real and imaginary parts of $\nu - \nu_0$, where $\nu$ is the frequency of a p-mode at $B_{\rm p} = 1$\,kG and $\nu_0$ is the frequency of the mode without a magnetic field.
The left panel compares the results of \cite[Cunha \& Gough (2000)]{cun00} for $\ell=1$ and 3 axisymmetric modes of the polytropic model having $(M,R)=(2M_\odot,2R_\odot)$ with the results of \cite[Bigot et al.~(2000)]{big00} for a $1.8\,M_\odot$ ZAMS model.  
The results of \cite[Cunha \& Gough (2000)]{cun00} show that  the frequency shift by the magnetic field increases with frequency, but at a certain frequency it jumps down and then starts increasing again.
This property, which is governed by a parameter, $\nu B_{\rm p}^{0.25}$ (for a polytrope of index 3), is confirmed by \cite[Saio \& Gautschy (2004a)]{sai00a}.
Although the calculations by \cite[Bigot et al.~(2000)]{big00} do not extend beyond the jump, the results agree with the results of \cite[Cunha \& Gough (2000)]{cun00} up to the jump.
  
The right panel of Fig.\,\ref{fig:comparison} compares the results of \cite[Bigot et al.~(2000)]{big00} for $\ell=0,2$ with those obtained by the method of \cite[Saio \& Gautschy (2004a)]{sai00a} for a $1.8\,M_\odot$ ZAMS model, which is similar to the model of \cite[Bigot et al.~(2000)]{big00}.
Obviously, both results agree well with each other.
Those comparisons show that the results from the three different methods agree well up to the jump, although some discrepancy in the amount of the jump is indicated by \cite[Saio (2008)]{sai08}.

\begin{figure}[b]
\begin{center}
 \includegraphics[width=0.55\textwidth]{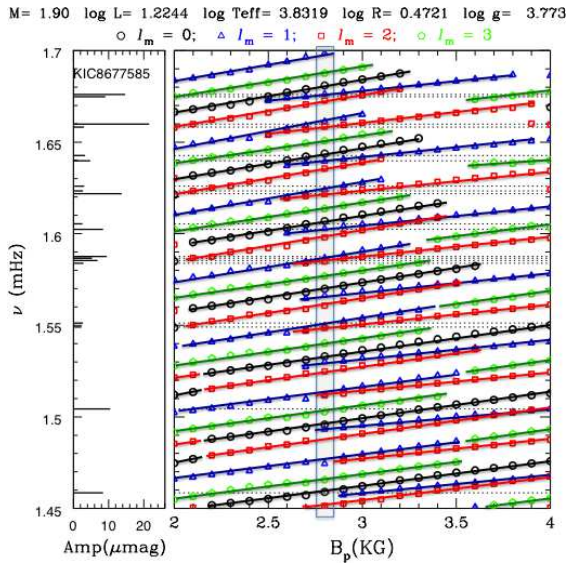} 
 \caption{\textit{Left} An amplitude spectrum of KIC 8677585 drawn using the data from \cite[Balona et al.~(2013)]{bal13}.  
 \textit{Right:} Frequencies of various modes calculated 
 by the method of \cite[Saio (2005)]{sai05} (a nonadiabatic extension of \cite[Saio \& Gautschy 2004]{sai04})
 are plotted as a function of $B_{\rm p}$ (magnetic strength at a pole).  The model with the parameters shown in the top line gives a best fit with the frequencies of KIC 8677585 at $B_{\rm p}=2.8$\,kG (the right panel of Fig.\,\ref{fig:echelle}). 
 Frequencies belonging to the same mode are connected by solid lines. Generally the frequency of a mode increases with $B_{\rm p}$, but at some $B_{\rm p}$ the effective latitudinal degree $\ell_m$ changes.}
   \label{fig:bpnu}
\end{center}
\end{figure}

\section{Comparison with KIC 8677585}

Among the three roAp stars found by the {\it Kepler} satellite, KIC 8677585 has the largest number of frequencies (\cite[Balona et al.~2013]{bal13}), which are more or less regularly spaced, but some frequencies form tight groups as seen in the left panel of Fig.\,\ref{fig:bpnu}. 
Such a group of high-order p modes  is one of the properties which can be explained by the effect of a magnetic field.

\begin{figure}[t]
\begin{center}
\includegraphics[width=0.51\textwidth]{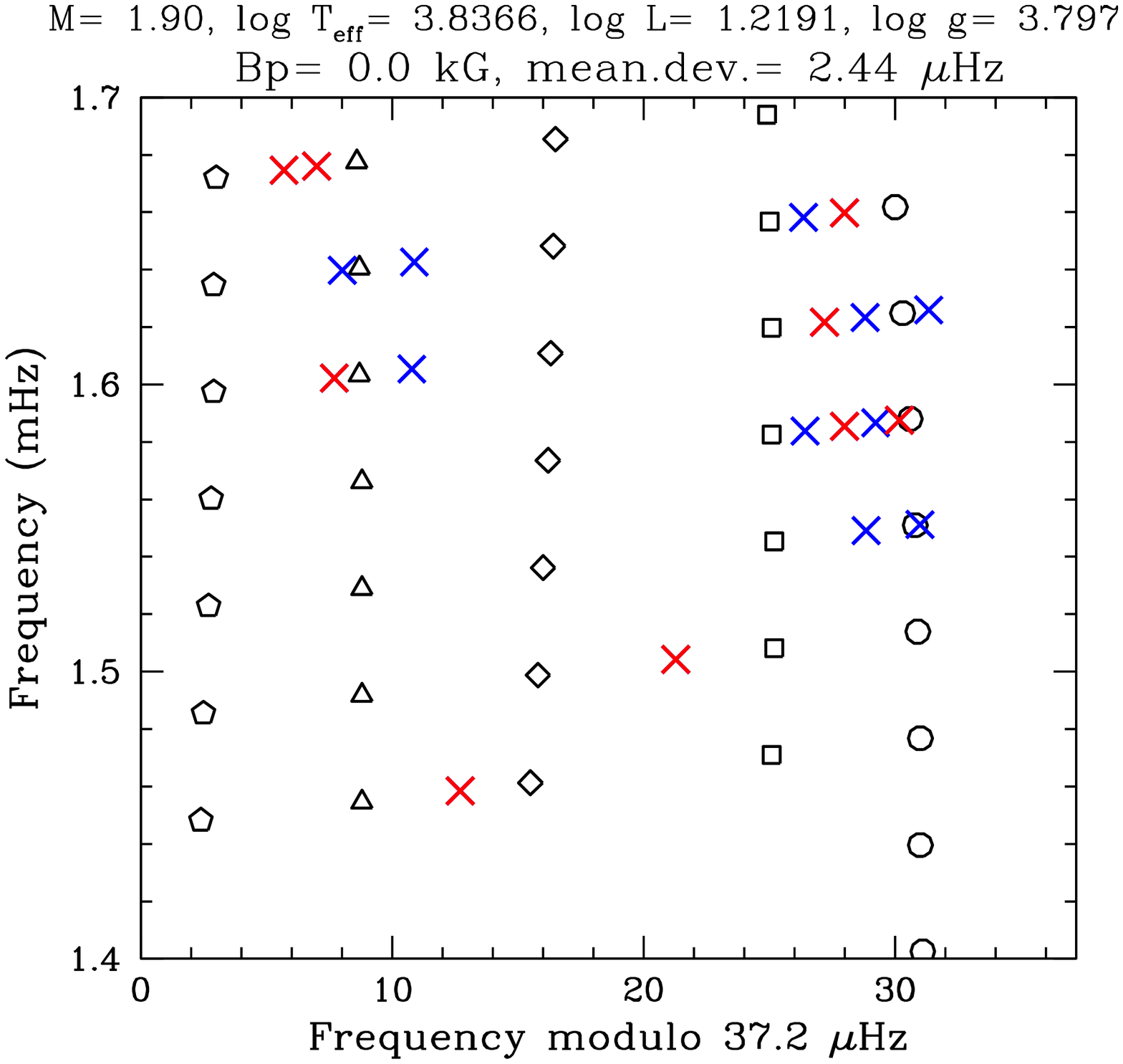}
\hspace{-0.04\textwidth}
\includegraphics[width=0.51\textwidth]{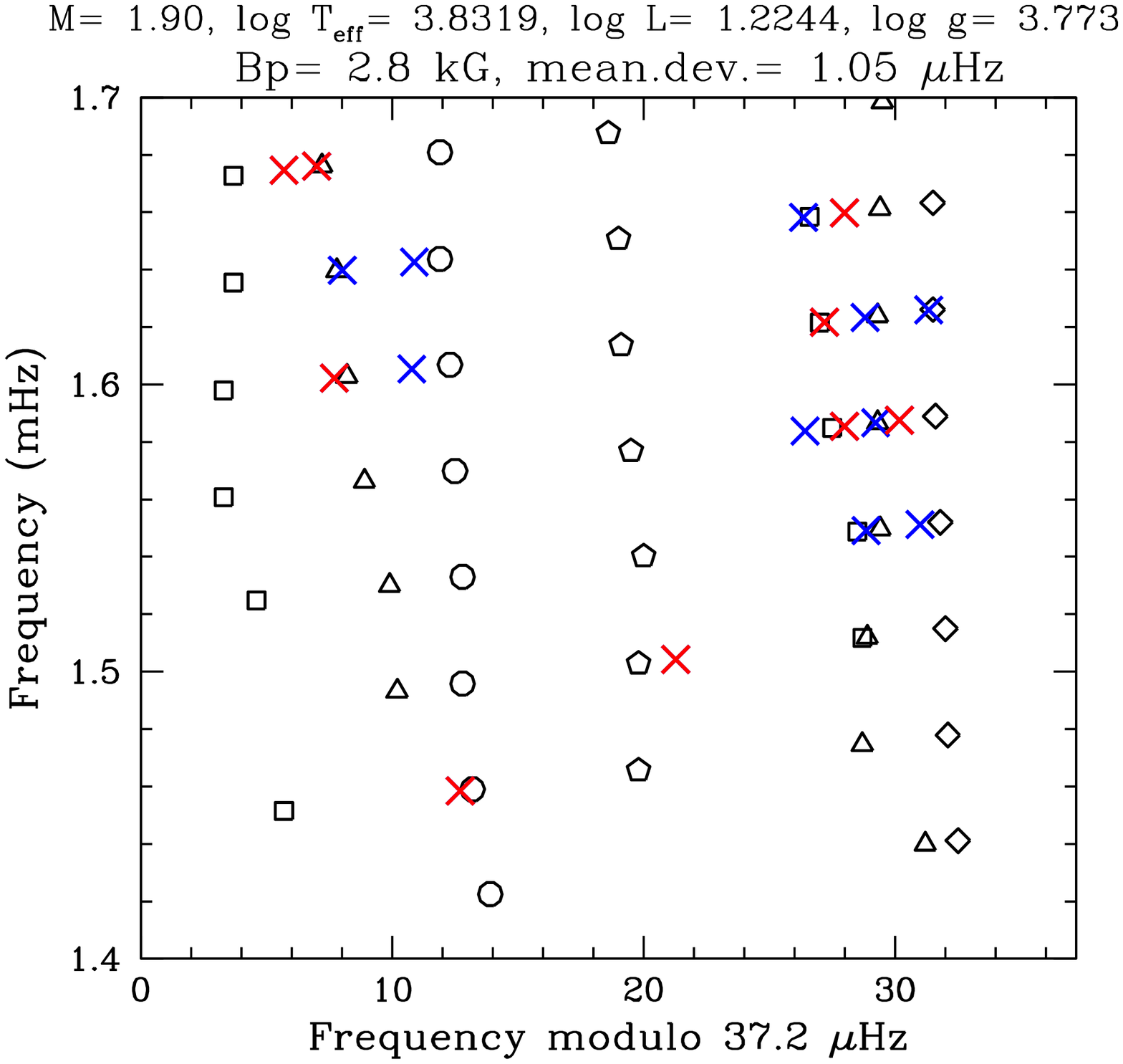}
\vspace{-0.8cm}
\caption{Oscillation frequencies of KIC 8677585 (crosses; \cite[Balona et al.~2013]{bal13}) from the {\it Kepler} satellite are compared with a model that does not include magnetic field (left panel), and with a model that does include the effect of a dipole magnetic field of 2.8 kG.}
\label{fig:echelle}
\end{center}
\end{figure}

As seen in the right panel of Fig.\,\ref{fig:bpnu}, frequency of a mode generally increases with $B_{\rm p}$.
But the rate of the increase (i.e., $d\nu/dB_{\rm p}$) depends slightly on the (effective) latitudinal degree ($\ell_m$).
Therefore, at some value of $B_{\rm p}$, frequencies of two (sometimes three) modes get very close.
Such a character can produce tight groupings seen in KIC 8677585.
An  echelle  diagram of the model produced at $B_{\rm p}=2.8$\,kG is compared with the observed frequencies in the right panel of Fig.\,\ref{fig:echelle}.
Compared with the left panel for a model with $B_{\rm p}=0$,  it is obvious that including the magnetic effect improves the fitting with the observed frequencies, 
indicating the theoretical prediction for the magnetic effect on the p-mode frequencies to be in the right direction.  However, we note that KIC 8677585 shows, in addition to the high-frequency oscillations, some low-frequency oscillations (34 -- 74\,$\mu$Hz; \cite[Balona et al.~2013]{bal13}). We do not understand how they are excited.

\section{Magnetic effects on the amplitude distribution at the surface}

Because of the Lorentz force, the latitudinal dependence of a pulsation mode  cannot be expressed by a single spherical harmonic. It changes gradually with the strength of the magnetic field.
Generally speaking, the amplitude on the surface is confined more strongly to
the polar regions, which corresponds to the increasing contribution from higher $\ell$ components (\cite[Saio \& Gautschy 2004a]{sai04a}).
This phenomenon was confirmed observationally by \cite[Kochukhov (2004)]{koc04}, who obtained the velocity amplitude distribution of the roAp star HR 3831 and found that the velocity amplitude of the ``dipole'' mode is more concentrated toward the magnetic axis than the dependence of the Legendre function $P_1(\cos\theta)$; i.e., there is a considerable contribution from $P_3(\cos\theta)$.

The modification of the amplitude distribution on the surface also affects the line-profile variations caused by the oscillations.
Contribution from higher $\ell$ components makes the ``wavelength'' of the profile variation shorter (\cite[Saio \& Gautschy 2004b]{sai04b}), in agreement with the observed ones (e.g., HR 3831 \cite[Kochukhov 2006]{koc06}).

\section{Effect of rotation -- Oblique pulsator model}
We discuss here the effects of rotation, which we  have disregarded so far.
Most of the roAp stars rotate slowly with an axis generally inclined to the magnetic axis (oblique rotator). 
To explain the observational property that the pulsation amplitude and phase modulate with the rotation phase, \cite[Kurtz (1982)]{kur82} invented the {\it oblique pulsator model}, in which the pulsation axis is aligned with the magnetic axis which is inclined to the rotation axis.
In this model, the alignment between the pulsation and the magnetic axes is assumed, but it is not strictly possible because the rotation effect, whose symmetry axis is inclined to the magnetic axis, modifies the pulsation.
\cite[Bigot \& Dziembowski (2002)]{big02} investigated this difficult problem and found that the pulsation axis of a nearly axisymmetric mode does not stay on one direction but draws an ellipse during a pulsation cycle.
Later, \cite[Bigot \& Kurtz (2011)]{big11} applied this modified oblique pulsator model to various roAp stars to fit observed amplitude and phase modulations better than the fits by the original model.

From a different point of view, \cite[Gough (2012)]{gou12} argued that even if the pulsation axis is once aligned to the magnetic axis, the alignment would be lost because it should precess due to the Coriolis force. To avoid the difficulty, he proposed that pulsation should be excited only when the axis is closely aligned to the magnetic axis.


\end{document}